# OBSERVATION OF THE ACCELERATED DIFFUSION IN NANOPARTICLES OF PARADIBROMOBENZENE /PARADIHLORBENZENE SOLID SOLUTION

**M.A.Korshunov**
*Institute of physics it. L. V.Kirenskogo of the Siberian separating of the Russian Academy of Sciences, 660036 Krasnoyarsk, Russia*
*e-mail: mkor@iph.krasn.ru*

P-dibromobenzene nanoparticles in a case of paradihlorbenzol molecules have been synthesized. Raman spectrums of these nanoparticles are measured. At diffusion of molecules of paradihlorbenzol in a p-dibromobenzene nanoparticle solid solution formed. Modifications in structure of nanoparticles at diffusion are reflected in Raman spectrums. Unlike a solid solution single crystal, in nanoparticles the accelerated diffusion, $D = 1.3\pm0.02\cdot10^{-11}$ cm$^2$/s, is observed at room temperature. Calculations show that the accelerated diffusion is caused by magnification of parametres of a lattice at reduction of sizes of nanoparticles.

In the molecular organic crystals diffusion flows past slow enough but because the tendency to use in practice nanocrystals was outlined it is necessary to know, how in such crystals diffusion flows. Migration of molecules can change physical properties and lead to violation of the given properties for use in practice.

For study of this problem p-dibromobenzene nanoparticles in a case of molecules of paradihlorbenzol by a size nearby 300nm have been synthesized. On the cooled glass substrate the thin film of paradihlorbenzol with the size about 300 nanometers has been created. After that, p-dibromobenzene nanoparticles have been sprayed onto the film. Then the film of paradihlorbenzol with the nanoparticles of a p-dibromobenzene arranged on it was evaporated. As melting point of paradihlorbenzol ($t_{mp}=56^0$) lower than of a p-dibromobenzene ($t_{mp}=86.8^0$), the paradihlorbenzol evaporated more promptly. It has led to that on a substrate there were p-dibromobenzene nanoparticles in a case of paradihlorbenzol molecules. The image of one of nanoparticles obtained by an electronic microscope is given in Fig. 1.

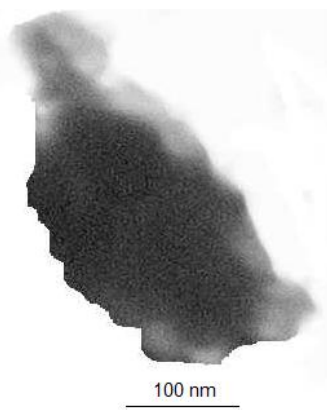

*Fig. 1. The image of a nanoparticle of a p-dibromobenzene in a paradihlorbenzol case gained on an electronic microscope*

Raman scattering allows to judge on spectrums of the lattice and intramolecular vibrations character of a disposition of molecules of an impurity among molecules of the basic crystal. In particular, whether on spectrums it is possible to spot solid solution as substitution was organized, introductions or impurity molecules are concentrated to boundaries between blocks of a lattice of a single crystal, or solid solution is a mechanical intermixture of substances. At initial manufacturing of nanoparticles on spectrums it is visible that molecules of paradihlorbenzol do not diffuse in a p-dibromobenzene particle as the spectrum is superposition of spectrums of a crystal of a p-dibromobenzene and paradihlorbenzol (Fig. 2). In Fig. 3(1) the spectrum of nanoparticles of solid solution after the lapse of 7 hours is given. The sample all time was at ambient temperature. Apparently it represents a spectrum of the mixed crystal with concentration of paradihlorbenzol of 30%. To gain a similar spectrum at diffusion study in a volume single crystal of a p-dibromobenzene apart 30 μm from edge where paradihlorbenzol is raised dust annealing of the sample is required at temperature 323K about 360 hours (the gained values of a diffusion constant for single crystal $D = 1.746 \pm 0.02 \cdot 10^{-12}$ cm$^2$/s at ambient temperature from calculations $D = 1.1 \pm 0.02 \cdot 10^{-13}$ cm$^2$/s) [1]. From experiment it is discovered that the nanoparticle diffusion constant at ambient temperature matters $D = 1.3 \pm 0.02 \cdot 10^{-11}$ cm$^2$/s thus time of diffusion for distance of a nanoparticle of 300 nanometers makes nearby 30 s. If we observed a surface diffusion in a spectrum we saw superposition, both a spectrum of the mixed crystal, and a spectrum of a pure p-dibromobenzene. In this case it is not observed. It is possible to judge concentration of components on a relation between intensities of intramolecular vibrations. Being besides grounded on a relation интенсивностей stretching vibrations C-Br and C-Cl and a spectrum of the lattice vibrations it is possible to judge what part of a crystal has organized a solid solution. In this case all nanoparticle represents solid solution.

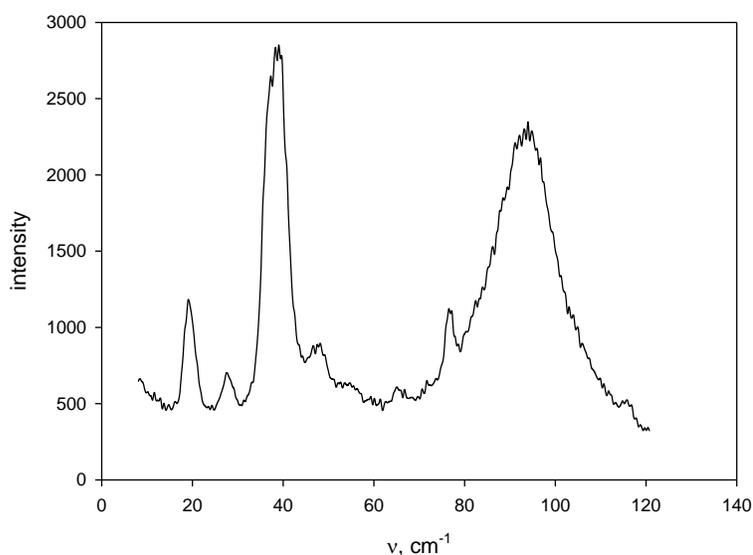

*Fig. 2. Spectrum Raman of a nanoparticle of a p-dibromobenzene in a paradihlorbenzol case.*

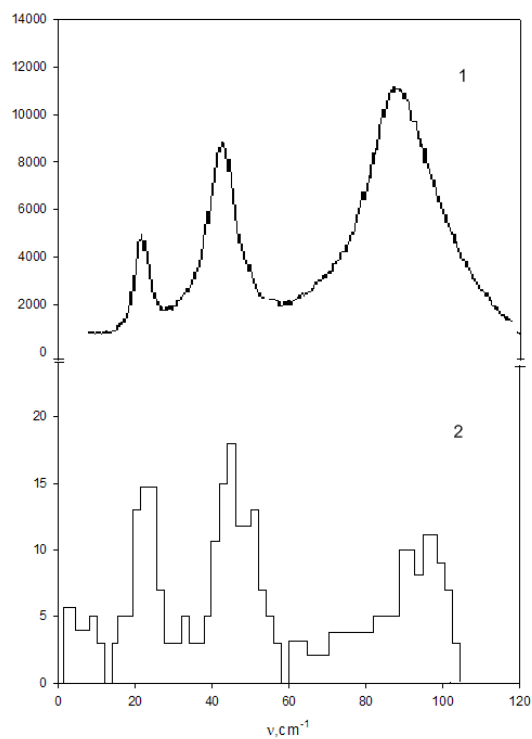

*Fig. 3. Spectrum Raman of a nanoparticle of solid solution paradibrombenzol/paradihlorbenzol*

Modeling of process of diffusion in solid solution have shown that diffusion rate is incremented at magnification of parameters of a lattice. The carried out calculations of a diffusion constant at migration of an impurity of paradihlorbenzol on a lattice of a nanoparticle of a p-dibromobenzene have shown that the value of a diffusion constant coinciding with the experimental value, is reached, when the solid solution lattice is incremented on a reference axis b and decreases on an axis z. It proves to be true modifications of frequency spectrums of nanoparticles and calculations of the lattice vibrations by Dyne's method [2]. The histogram of the calculated spectrum of the lattice vibrations of solid solution at the account of magnification of parameters of a lattice is given in Fig. 3(2). Concentration of molecules of paradihlorbenzol thus was 30%.

Thus, at reduction of sizes of nanoparticles of solid solutions paradibrombenzola/paradihlorbenzola because of the magnification of parametres of a lattice the diffusion rate in nanoparticles increases.